\documentclass[aps,prd,superscriptaddress,twoside,twocolumn,nofootinbib,showpacs,floatfix]{revtex4-1}
\usepackage{amsmath,amssymb}
\usepackage{graphicx,bm,braket}
\usepackage{subfigure}
\usepackage[colorlinks,linkcolor=blue]{hyperref}
\usepackage{color}
\usepackage{newtxtext, newtxmath}
\pdfpagewidth=\paperwidth
\pdfpageheight=\paperheight
\allowdisplaybreaks
\usepackage{epstopdf}

\newcommand*{\hc}{\text{H.\,c.}}

\begin{document}

\title{\boldmath Effects of hadronic molecule $N(2080)3/2^-$ on $K^{*+}\Lambda$ photoproduction}

\author{Wen-Ya Tian}
\affiliation{School of Nuclear Science and Technology, University of Chinese Academy of Sciences, Beijing 101408, China}

\author{Neng-Chang Wei}
\affiliation{School of Physics, Henan Normal University, Henan 453007, China}

\author{Yu-Fei Wang}
\affiliation{School of Nuclear Science and Technology, University of Chinese Academy of Sciences, Beijing 101408, China}

\author{Fei Huang}
\email{huangfei@ucas.ac.cn}
\affiliation{School of Nuclear Science and Technology, University of Chinese Academy of Sciences, Beijing 101408, China}

\author{Bing-Song Zou}
\email{zoubs@mail.tsinghua.edu.cn}
\affiliation{Department of Physics, Tsinghua University, Beijing 100084, China}
\affiliation{CAS Key Laboratory of Theoretical Physics, Institute of Theoretical Physics, Chinese Academy of Sciences, Beijing 100190, China}
\affiliation{School of Physical Sciences, University of Chinese Academy of Sciences, Beijing 100049, China}
\affiliation{Southern Center for Nuclear-Science Theory (SCNT), Institute of Modern Physics, Chinese Academy of Sciences, Huizhou 516000, China}

\date{\today}

\begin{abstract}
In our previous work [Phys. Rev. C {\bf 101}, 014003 (2020)], we have analyzed all available data on differential cross sections and spin density matrix elements for the $\gamma p \to K^{\ast +} \Lambda$ reaction using an effective Lagrangian approach. There, the $t$-channel $K$, $K^*$, and $\kappa$ exchanges, the $u$-channel $\Lambda$, $\Sigma$, and $\Sigma^*$ exchanges, the $s$-channel $N$, $N(2060)5/2^-$, and $N(2000)5/2^+$ exchanges, and the interaction current were taken into account in constructing the reaction amplitudes. It was found that, overall, the agreement of the theoretical results with the corresponding data is fairly good. However, noticeable discrepancies between the model results and the data on spin density matrix elements $\rho_{00}$ are still observed in the center-of-mass energy region around $W \approx 2.08$ GeV, indicating the need for additional reaction mechanisms. On the other hand, the hidden-strange pentaquark-like state $N(2080)3/2^-$, which is proposed to be a $K^*\Sigma$ molecular state as the strange partner of $P_c(4457)$ hadronic molecule, has been found to play quite active roles in reproducing the data for $K^{*+}\Sigma^0$,  $K^{*0}\Sigma^+$, and $\phi p$ photoproduction reactions. Based on these observations, in the present work, we reanalyze the data for the $\gamma p \to K^{\ast +} \Lambda$ reaction by incorporating the pentaquark-like state $N(2080)3/2^-$ in our previously proposed model. The results show that with the inclusion of the contributions of $N(2080)3/2^-$, the quality of the theoretical description of the $\gamma p \to K^{*+}\Lambda$ data can be considerably improved. In particular, the data on spin density matrix elements $\rho_{00}$ and ${\rm Re}\,\rho_{10}$ in the near-threshold energy region are significantly better reproduced. This achievement indicates that the hypothesis of the $N(2080)3/2^-$ molecular state is compatible with the available data for the $\gamma p \to K^{*+}\Lambda$ reaction. The reaction mechanism of $\gamma p \to K^{*+}\Lambda$ is analyzed and the $N(2080)3/2^-$, $N(2060)5/2^-$, $N(2000)5/2^+$, and $K$ exchanges are found to provide dominant contributions.
\end{abstract}



\maketitle

\section{Introduction}   \label{Sec:intro}

Quantum chromodynamics (QCD) is the foundational theory of the strong interactions among quarks and possesses two main features: asymptotic freedom and color confinement. The simplest compositions for quarks to form color singlet hadrons are triplets of quarks, $qqq$, which constitute baryons, and quark-antiquark pairs, $q\bar{q}$, which form mesons. These {\it normal} hadrons have been extensively studied and observed in numerous experiments, forming the basis of our current understanding of particle physics. However, QCD does not limit itself to these traditional configurations. The theory also allows for the existence of {\it exotic} hadronic states composed of more than the typical number of quarks, such as baryons made of five quarks (pentaquark states) or six quarks (hexaquark states), as well as mesons made of two quarks and two antiquarks (tetraquark states). Since the early 2000s, experimental collaborations such as LHCb, CMS, Belle, BESIII, Fermilab, and others have provided compelling evidences for the existences of a series of tetraquark mesons and pentaquark baryons \cite{Belle:2003nnu,LHCb:2013kgk,CMS:2021znk,Cotugno:2009ys,Chen:2022asf,LHCb:2015yax,LHCb:2019kea}. These discoveries challenge the conventional understanding of quark combinations and their interactions, and ignite ongoing debates regarding the internal structures and quantum numbers of these {\it exotic} hadrons.

In the pentaquark sector, the first compelling evidence was reported by the LHCb Collaboration, confirming the existence of the $P_c^+(4312)$, $P_c^+(4380)$, $P_c^+(4440)$, and $P_c^+(4457)$ states \cite{LHCb:2015yax,LHCb:2019kea}. These states cannot be accommodated as three-quark excitation states, as their excitation energies exceed $3$ GeV, significantly higher than those of the traditional $N^\ast$ and $\Delta^\ast$ states, which are typically less than $2$ GeV. Instead, they are interpreted as hidden-charm pentaquark states or baryon-meson states as predicted in Refs.~\cite{Wu:2010jy,Wang:2011rga,Yang:2011wz,Wu:2012md,Xiao:2013yca}. In literature, one of the most plausible explanations for the $P_c^+(4380)$ and $P_c^+(4457)$ states is that they are hadronic molecules composed of ${\bar D}\Sigma_c^\ast$ and ${\bar D}^\ast \Sigma_c$, respectively, as their reported masses are approximately $2$ MeV below the thresholds of ${\bar D}\Sigma_c^\ast$ ($4382$ MeV) and ${\bar D}^\ast \Sigma_c$ ($4459$ MeV). See, for example, a recent study on this topic~\cite{Wang:2025ecf}. It is worth noting that the theoretically predicted $\bar{D}\Sigma^*_c$ hadronic molecule is more narrow compared with the $P_c^+(4390)$ reported by LHCb in 2015, which disappeared in later experimental analyses \cite{Du:2019pij,Liu:2019tjn}.  

In the light quark sector, the $N(1875)3/2^-$ and $N(2080)3/2^-$ states have been proposed to be the strange partners of the $P_c^+(4380)$ and $P_c^+(4457)$ molecular states \cite{Lin:2018kcc,He:2017aps,Zou:2018uji}. They are hypothesized to be hadronic molecular states composed of $K\Sigma^\ast$ and $K^\ast \Sigma$ whose thresholds are at $1880$ MeV and $2086$ MeV, respectively, and are referred to as $P_s$ hidden-strange states. Recently, the roles of these $P_s$ states have been investigated in the $K^{*+}\Sigma^0$, $K^{*0}\Sigma^+$, and $\phi p$ photoproduction reactions that involve $s\bar{s}$ content in the final states \cite{Ben:2023uev,Wu:2023ywu,Shim:2024myp}. In these studies, a phase factor was introduced to model the molecular nature of $P_s$, and it was found that these $P_s$ states are rather helpful in explaining the data for these reactions. Reference \cite{Shi:2025mvl} reports that the $3/2^-$-nucleon resonance around $2080$ MeV is strongly coupled with the $K^*\Sigma$ final state, supporting the molecular picture of $N(2080)3/2^-$.

In the present work, we focus on the $\gamma p \to K^{*+}\Lambda$ reaction, with the primary objective of assessing the compatibility of the hadronic molecular picture of the $P_s$ states with the available data for this process. Different to the aforementioned studies of Refs.~\cite{Ben:2023uev,Wu:2023ywu,Shim:2024myp} where the molecular nature of the $P_s$ states was modeled by additional phase factors attached to the tree-level reaction amplitudes, here we directly calculate the full triangle loop diagrams for couplings of the $P_s$ states to the initial $\gamma p$ and final $K^{*+}\Lambda$ states. Though after the main analyses, we will also catch a glimpse of the parametrization using phase factors. This exploration is expected to provide further insights into the nature of these exotic $P_s$ states and their role in the dynamics of photoproduction reactions.

\begin{figure}[tbp]
    \includegraphics[width=\columnwidth]{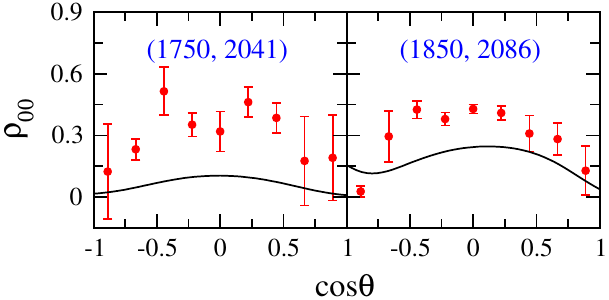}
    \caption{The model results for the spin density matrix elements $\rho_{00}$ for $\gamma p\rightarrow K^{*+}\Lambda$ as a function of $\cos \theta$ in the center-of-mass frame at $W = 2041$ MeV and $2086$ MeV from Ref.~\cite{Wei:2020fmh} compared with the corresponding experimental data \cite{ CLAS:2017sgi}. The numbers in  parentheses on the left denote the photon laboratory incident energies, in MeV.}
    \label{fig_wei}
\end{figure}

Actually, in our previous work \cite{Wei:2020fmh}, we have already analyzed all available data \cite{CLAS:2013qgi,CLAS:2017sgi} on differential cross sections and spin density matrix elements for the $\gamma p \to K^{*+}\Lambda$ reaction using a tree-level effective Lagrangian approach. There, the reaction amplitudes were constructed by including the $t$-channel $K$, $K^*$, and $\kappa$ exchanges, the $u$-channel $\Lambda$, $\Sigma$, and $\Sigma^*$ exchanges, the $s$-channel $N$, $N(2060)5/2^-$, and $N(2000)5/2^+$ exchanges, and the interaction current. It was found that the built model can provide an overall good description of all the available data, with the dominant contributions arising from the $t$-channel $K$ exchange and the $s$-channel exchanges of $N(2060)5/2^-$ and $N(2000)5/2^+$. In spite of these successes, noticeable discrepancies between the model results and the data for the spin density matrix elements $\rho_{00}$ were still observed in the center-of-mass energy region around $W \approx 2.08$ GeV, as illustrated in Fig.~\ref{fig_wei}, where one sees clearly that the data on $\rho_{00}$ around $W \approx 2.08$ GeV are significantly underestimated by the model. This discrepancy suggests the need for additional reaction mechanisms within our theoretical framework to better describe the data in this energy range. A natural candidate for such a mechanism is the exchange of $N(2080)3/2^-$, a $K^\ast \Sigma$ molecular state proposed as the strange partner of the $P_c^+(4457)$ \cite{Lin:2018kcc,He:2017aps,Zou:2018uji}.

In the present work, we reanalyze all available data for the $\gamma p \to K^{*+}\Lambda$ reaction by further considering the contributions from the $N(2080)3/2^-$ molecular state based on our previous work of Ref.~\cite{Wei:2020fmh}. We anticipate that after the inclusion of the $N(2080)3/2^-$ exchange, the theoretical description of the $\gamma p \to K^{*+}\Lambda$ data in our model can be much improved, and thus, a more reliable understanding of the reaction mechanisms especially the resonance parameters and their contributions can be achieved. If really so, this, in turn, would provide support for the hadronic molecular interpretation of the $N(2080)3/2^-$ state, demonstrating its compatibility with the available data for the $\gamma p \to K^{*+}\Lambda$ reaction.

\begin{figure}[tp]
    \subfigure[]{
    \includegraphics[width=0.4\columnwidth]{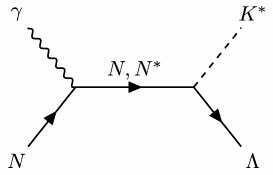}}  {\hglue 0.5cm}
    \subfigure[]{
    \includegraphics[width=0.4\columnwidth]{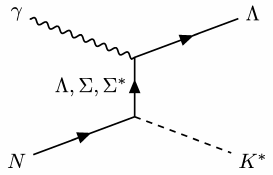}} \\[6pt]
    \subfigure[]{
    \includegraphics[width=0.4\columnwidth]{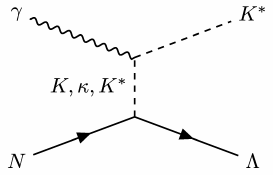}} {\hglue 0.5cm}
    \subfigure[]{
    \includegraphics[width=0.4\columnwidth]{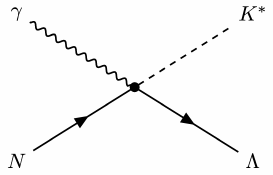}}
    \caption{Generic structure of the photoproduction amplitude for $\gamma N\to K^{*}\Lambda$. Time proceeds from left to right. (a) $s$ channel, (b) $u$ channel, (c) $t$ channel, and (d) Interaction current.}
    \label{FIG:feymans}
\end{figure}

The paper is organized as follows. In Sec.~\ref{Sec:formalism}, we briefly introduce the theoretical framework of our calculation. The results of the model, along with relevant discussions, are presented in Sec.~\ref{Sec:results}. Finally, a brief summary and conclusions are given in Sec.~\ref{sec:summary}.

\section{Formalism}  \label{Sec:formalism}

\begin{figure}[tbp]
    \subfigure[]{
    \includegraphics[width=0.45\columnwidth]{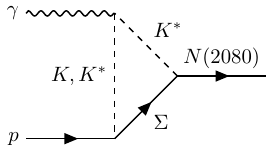}}   {\hglue 0.5cm}
    \subfigure[]{
    \includegraphics[width=0.45\columnwidth]{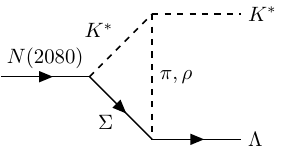}} \\[6pt]
    \subfigure[]{
    \includegraphics[width=0.45\columnwidth]{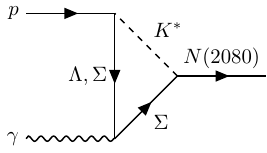}}  {\hglue 0.5cm}
    \subfigure[]{
    \includegraphics[width=0.45\columnwidth]{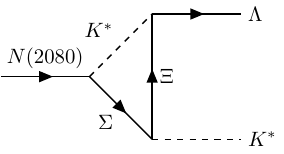}}
    \caption{Couplings of $N(2080)3/2^-$ as $K^*\Sigma$ molecule to $K^{*}\Lambda$ and $\gamma N$.}
    \label{fig:loop}
\end{figure}

As developed in Ref.~\cite{Wei:2020fmh} and schematically depicted in Fig.~\ref{FIG:feymans}, the reaction amplitudes for $\gamma N\to K^{*}\Lambda$ in our effective Lagrangian model are constructed by including the $s$-channel $N$ and $N^*$ exchanges, the $t$-channel $K$, $K^*$, and $\kappa$ exchanges, the $u$-channel $\Lambda$, $\Sigma$, and $\Sigma^*$ exchanges, as well as the interaction current. In Ref.~\cite{Wei:2020fmh}, it was found that the exchanges of the $N(2060)5/2^-$ and $N(2000)5/2^+$ resonances in the $s$ channel are needed to explain the near-threshold structures exhibited in the angular distribution data and are essential for providing an overall description of the available data on both the differential cross sections and the spin density matrix elements for $\gamma p\to K^{*+}\Lambda$. As the data on spin density matrix elements $\rho_{00}$ near $W\approx 2.08$ GeV are much underestimated in Ref.~\cite{Wei:2020fmh}, as shown in Fig.~\ref{fig_wei}, in this work, we further include in $s$ channel the exchange of the pentaquark-like state $N(2080)3/2^-$ and reanalyze all available data for $\gamma p \to K^{*+}\Lambda$ within the theoretical framework proposed in our previous work of Ref.~\cite{Wei:2020fmh}.

The full reaction amplitude ${\cal M}$ for $\gamma N\to K^{*}\Lambda$ can be written as
\begin{eqnarray}
{\cal M} = {\cal M}_s + {\cal M}_t + {\cal M}_u + {\cal M}_{\rm int},  \label{eq:amplitude}
\end{eqnarray}
where ${\cal M}_s$, ${\cal M}_t$, and ${\cal M}_u$ denote the amplitudes calculated from the $s$-, $t$-, and $u$-channel reaction mechanisms, respectively. ${\cal M}_{\rm int}$ denotes the term calculated from the interaction current which is introduced to ensure the gauge invariance of the full amplitude, ${\cal M}$, of the $K^{*}\Lambda$ photoproduction \cite{Haberzettl:1997,Haberzettl:2006,Huang:2012,Huang:2013}. 

For the $s$-channel exchange of $N(2080)3/2^-$ which is proposed to be a hadronic molecular state of $S$-wave $K^*\Sigma$, its couplings to $K^{*}\Lambda$ and $\gamma N$ are obtained by directly calculating the triangle diagrams as shown in Fig.~\ref{fig:loop}. Note again that we start with the full triangle diagrams rather than simplified parametrizations by phase factors in Refs.~\cite{Ben:2023uev,Wu:2023ywu,Shim:2024myp}. The relevant effective Lagrangians contain:
\begin{align}
\mathcal{L}_{R\Sigma K^*} =&~  g_{R\Sigma K^*}\bar{R}_\mu K^{*\mu}\Sigma + \hc, \\[6pt]
{\cal L}_{\gamma K^* K^*} =& -i e \left({K^{*+}}^\nu K^{*-}_{\mu\nu} -{K^{*-}}^\nu K^{*+}_{\mu\nu}\right) A^\mu, \\[6pt]
{\cal L}_{\gamma K{K^*}} =&~ e\frac{g_{\gamma K{K^*}}}{M_K}\varepsilon^{\alpha \mu \lambda \nu}\left(\partial_\alpha A_\mu\right)\left(\partial_\lambda K\right)K^*_\nu, \label{Lag:gKKst} \\[6pt]
\mathcal{L}_{\Sigma\Sigma\gamma} =& -e\bar{\Sigma}\left[\left(\hat{e}\gamma^\mu-\frac{\hat{\kappa}_{\Sigma}}{2M_N}\sigma^{\mu\nu}\partial_\nu\right)A_\mu\right]\Sigma, \\[6pt]
{\cal L}_{\Sigma \Lambda \gamma} =&~ e\frac{\kappa_{\Sigma \Lambda}}{2M_N}\bar{\Lambda}\sigma^{\mu \nu}\left(\partial_\nu A_\mu\right)\Sigma^0 + \hc,  \\[6pt]
{\cal L}_{\Sigma NK} =& - \frac{g_{\Sigma NK}}{2M_N} \bar{\Sigma}\gamma^5 \gamma^\mu \left(\partial_\mu K \right) N + \hc,  \label{eq:L_LNK}    \\[6pt]
{\cal L}_{\Sigma N {K^*}} =& - g_{\Sigma N {K^*}}\bar{\Sigma}\left[\left(\gamma^\mu-\frac{\kappa_{\Sigma N {K^*}}}{2M_N}\sigma^{\mu \nu}\partial_\nu\right)K^*_\mu\right] N + \hc, \\[6pt]
{\cal L}_{\Lambda N {K^*}} =& - g_{\Lambda N {K^*}}\bar{\Lambda}\left[\left(\gamma^\mu-\frac{\kappa_{\Lambda N {K^*}}}{2M_N}\sigma^{\mu \nu}\partial_\nu\right)K^*_\mu\right] N + \hc, \\[6pt]
\mathcal{L}_{\Sigma\Lambda\rho} =& -g_{\Sigma\Lambda\rho}\bar{\Lambda}\left[\left(\gamma^\mu-\frac{\kappa_{\Sigma\Lambda\rho}}{2M_N}\sigma^{\mu\nu}\partial_\nu\right)\rho_\mu\right]\Sigma + \hc, \\[6pt]
\mathcal{L}_{\Sigma\Lambda\pi} =& -\frac{g_{\Sigma\Lambda\pi}}{2M_N}\bar{\Lambda}\gamma^5\gamma^\mu \left(\partial_\mu\pi\right)  \Sigma + \hc, \\[6pt]
\mathcal{L}_{\Xi\Lambda K^*} =& -g_{\Xi\Lambda K^*}\bar{\Xi}\left[\left(\gamma^\mu-\frac{\kappa_{\Xi\Lambda K^*}}{2M_N}\sigma^{\mu\nu}\partial_\nu\right)\bar{K}^{*}_\mu\right]\Lambda + \hc, \\[6pt]
\mathcal{L}_{\Xi\Sigma K^*} =& -g_{\Xi\Sigma K^*}\bar{\Xi}\left[\left(\gamma^\mu-\frac{\kappa_{\Xi\Sigma K^*}}{2M_N}\sigma^{\mu\nu}\partial_\nu\right)\bar{K}^*_\mu\right]\Sigma + \hc,\\[6pt]
\mathcal{L}_{K^*K^*\pi} =&~ \frac{g_{K^*K^*\pi}}{M_\omega}\epsilon^{\mu\nu\alpha\beta}\left(\partial_\mu K^{*}_\nu\right) \left(\partial_\alpha K^*_\beta\right) \pi,\\[6pt] 
\mathcal{L}_{K^*K^*\rho} =& -ig_{K^*K^*\rho}\left(K^{*\mu\dagger} \rho^{\nu}K^{*}_{\mu\nu}-K^{*\dagger}_{\mu\nu}\rho^{\nu}K^{*\mu}\right. \nonumber \\[3pt]
   & \left. + \, K^{*\nu\dagger}\rho_{\mu\nu}K^{*\mu}\right), 
\end{align}
where the factors in isospin spaces are omitted, $e$ is the elementary charge unit, and $\hat{e}$ is the charge operator. The operator $\hat{\kappa}_\Sigma$ is given by $\hat{\kappa}_\Sigma\equiv\kappa_{\Sigma^+}(1+\hat{e})/2+\kappa_{\Sigma^-}(1-\hat{e})/2$. The field tensor for vector meson is defined as $V^{\mu\nu}\equiv \partial^\mu V^\nu - \partial^\nu V^\mu$. 

For the coupling of $N(2080)3/2^-$ to $K^*\Sigma$, due to the interpretation of $N(2080)3/2^-$ as a $K^*\Sigma$ molecule, we follow Ref.~\cite{Ben:2023uev} to assign the value based on Weinberg's compositeness criterion \cite{Weinberg:1965}: 
\begin{equation}
g^2_{R\Sigma K^*}=\frac{4\pi}{4m_Rm_\Sigma}\frac{\left(m_{K^*}+m_\Sigma\right)^{5/2}}{\left(m_{K^*}m_{\Sigma}\right)^{1/2}}\sqrt{32\epsilon},
\end{equation}
where $\epsilon\equiv m_{K^*}+m_\Sigma-m_R$ is the binding energy; $m_R, m_\Sigma$, and $m_{K^*}$ are the masses of $N(2080)3/2^-$, $\Sigma$, and $K^*$, respectively. Then the resulting value is $g_{R\Sigma K^*}=1.72$.

The electromagnetic couplings needed for the calculation of triangle vertices in Fig.~\ref{fig:loop} are determined as follows. The $\gamma KK^{*}$ coupling is determined by $K^{*}\rightarrow K\gamma$ decay, yielding $g_{\gamma K^\pm K^{*\pm}}=0.41$ and $g_{\gamma K^0K^{*0}}=-0.63$. The sign difference stems from the difference in the electric charges of the constituent quarks. The anomalous magnetic moments $\kappa_{\Sigma^+}=1.46$, $\kappa_{\Sigma^-}=-0.16$, and $\kappa_{\Sigma\Lambda}=-1.61$. 

Most of the hadronic couplings needed for the calculation of triangle vertices in Fig.~\ref{fig:loop} are determined based on SU(3) flavor symmetry \cite{Huang:2012,Ronchen:2012eg}:
\begin{align}
g_{\Sigma NK} =&~ \frac{1}{5} g_{NN\pi} = 2.69, \\[6pt]
g_{\Sigma NK^*} =& -\frac{1}{2} g_{NN\omega} + \frac{1}{2} g_{NN\rho} = -4.26, \\[6pt]
\kappa_{\Sigma N K^*} =&~ \frac{1}{2}\frac{f_{NN\rho}}{g_{\Sigma N K^*}} = -2.33,\\[6pt]
g_{\Lambda NK^*} =& -\frac{1}{2\sqrt{3}} g_{NN\omega} - \frac{\sqrt{3}}{2} g_{NN\rho} = -6.21, \\[6pt]
\kappa_{\Lambda N K^*} =& -\frac{\sqrt{3}}{2}\frac{f_{NN\rho}}{g_{\Lambda N K^*}} = 2.76,\\[6pt]
g_{\Sigma\Lambda\rho} =& -\frac{1}{2\sqrt{3}}g_{NN\omega}+\frac{\sqrt{3}}{2}g_{NN\rho}=-0.58,\\[6pt]
\kappa_{\Sigma\Lambda\rho} =&~ \frac{\sqrt{3}}{2}\frac{f_{NN\rho}}{g_{\Sigma\Lambda\rho}}=-29.59,\\[6pt]
g_{\Sigma \Lambda \pi} =&~ \frac{2\sqrt{3}}{5} g_{NN\pi} = 9.32, \\[6pt]
g_{\Xi\Lambda K^*} =&~ \frac{1}{\sqrt{3}}g_{NN\omega}=6.79,\\[6pt]
\kappa_{\Xi\Lambda K^*} =&~ \frac{1}{\sqrt{3}}\frac{f_{NN\omega}}{g_{\Xi\Lambda K^*}}=0,\\[6pt]
g_{\Xi\Sigma K^*} =& -g_{NN\rho}=-3.25, \\[6pt]
\kappa_{\Xi\Sigma K^*} =&- \frac{f_{NN\rho}}{g_{\Xi\Sigma K^*}}=6.10.
\end{align}
Here the empirical values $g_{NN\pi}=13.46$, $g_{NN\rho}=3.25$, $g_{NN\omega}=11.76$, $\kappa_{NN\rho}=f_{NN\rho}/g_{NN\rho}=6.1$, and $f_{NN\omega}=0$ from Refs.~\cite{Huang:2012,Ronchen:2012eg} are used. The hadronic couplings $g_{K^*K^*\pi} =-5.0$ and $g_{K^*K^*\rho} = 3.02$ are taken from Ref.~\cite{Lin:2018kcc} (The value $g_{K^*K^*\pi} =-5.0$ differs from that in Ref.~\cite{Lin:2018kcc} because of the different forms of the Lagrangian ${\cal L}_{K^*K^*\pi}$).

To account for the weakly bound molecular nature of the $N(2080)3/2^-$ resonance, a form factor is introduced in each triangle loop of Fig.~\ref{fig:loop} to suppress short-distance contributions, which are not expected to dominate in such a shallow binding regime. The adopted form factor reads 
\begin{equation}
f(q^2) = \frac{\Lambda_{M,B}^4}{\Lambda_{M,B}^4 + (q^2-m^2)^2},
\end{equation}
with $m$ and $q$ being the mass and four-momentum of the exchanged particle in the triangle loop diagram. The cutoff $\Lambda_M$ and $\Lambda_B$ (for meson and baryon exchanges, respectively) are determined by fitting the experimental data.  

The contributions of interaction terms to ${\cal M}_s$, ${\cal M}_t$, and ${\cal M}_u$ in Eq.~\eqref{eq:amplitude} can be obtained straightforwardly by calculating the corresponding Feynman diagrams. The necessary vertex Lagrangians, resonance propagators, and phenomenological form factors required for these calculations, as well as the construction of the interaction current $M_{\rm int}$ and the formulas for the differential cross section $d\sigma/d\Omega$ and the spin density matrix elements $\rho$, have been thoroughly introduced and detailed in Refs.~\cite{Wei:2020fmh,Wang:2017tpe}. For the sake of brevity and simplicity, we do not repeat them here.

\begin{table}[tbp]
    \caption{Fitted values for parameters of $N(2000)5/2^+$ and $N(2060)5/2^+$. }
    \label{tab:para1}
    \begin{tabular*}{\columnwidth}{@{\extracolsep\fill}lrr}
        \hline\hline
           & $N(2000)5/2^+$ & $N(2060)5/2^-$ \\ \hline
            $M_R$ [MeV] & $2009 \pm 1$ & $2035 \pm 1$ \\
            $\Gamma_R$ [MeV] & $400 \pm 1$ & $70 \pm 4$ \\
            $\Lambda_R$ [MeV] & $1080 \pm 8$ & $1318 \pm 10$ \\
            $\sqrt{\beta_{\Lambda K^*}}A_{1/2}~[10^{-3}~\text{GeV}^{-1/2}]$ & $0.11 \pm 0.01$ & $0.44 \pm 0.01$ \\
            $\sqrt{\beta_{\Lambda K^*}}A_{3/2}~[10^{-3}~\text{GeV}^{-1/2}]$ & $0.37 \pm 0.01$ & $-0.94 \pm 0.01$ \\
            $g^{(2)}_{R\Lambda K^*}/g^{(1)}_{R\Lambda K^*}$ & $-2.29 \pm 0.04$ & $0.35 \pm 0.25$\\
            $g^{(3)}_{R\Lambda K^*}/g^{(1)}_{R\Lambda K^*}$ & $-0.97 \pm 0.05$ & $5.54 \pm 0.54$ \\
        \hline\hline
    \end{tabular*}
\end{table}

\begin{table}[tbp]
    \caption{Fitted values for $K$, $\Sigma^*$, and $N(2080)3/2^-$ exchanges.}
    \label{tab:para2}
    \begin{tabular*}{\columnwidth}{@{\extracolsep\fill}lr}
    \hline\hline
        $\Lambda_K$ [MeV] & $945 \pm 4$ \\
        $g^{(1)}_{\gamma\Lambda\Sigma^*}$ & $-1.29 \pm 0.34$ \\
        $\Lambda_{M}$ [MeV] & $837 \pm 7$ \\
        $\Lambda_{B}$ [MeV] & $733 \pm 9$ \\
     \hline\hline
    \end{tabular*}
\end{table}

\begin{table}[tbp]
    \caption{Evaluated $\chi_i^2/N_i$ in the present work and Ref.~\cite{Wei:2020fmh} for a given type of observable specified by the index $i =d\sigma$ (differential cross section), $\rho_{00}$, ${\rm Re}\,\rho_{10}$, and $\rho_{1-1}$, with $N_i$ being the corresponding number of data points considered. The numbers in the brackets denote the corresponding values for the data in the energy range $W\le 2217$ MeV. The last row corresponds to the global $\chi^2/N$, where $N$ is the total number of data points including all the types of observables considered. }
\begin{tabular*}{\columnwidth}{@{\extracolsep\fill}lcc}
    \hline\hline
  &   Ref.~\cite{Wei:2020fmh}  &  The present work  \\ \hline
$\chi_{d\sigma}^2/N_{d\sigma}$  &  $2.79~(1.23)$  &  $2.94~(1.29)$    \\
$\chi_{\rho_{00}}^2/N_{\rho_{00}}$  &  $6.26~(9.03)$   &   $5.73~(7.70)$   \\
$\chi_{{\rm Re}\,\rho_{10}}^2/N_{{\rm Re}\,\rho_{10}}$  &  $5.28~(2.35)$   &  $5.98~(1.51)$   \\
$\chi_{\rho_{1-1}}^2/N_{\rho_{1-1}}$ &   $6.91~(5.52)$  &  $5.65~(5.26)$   \\
$\chi^2/N$  &   $5.27~(4.53)$     &  $5.04~(3.94)$    \\
  \hline\hline
\end{tabular*}
\label{table_chi}
\end{table}

\begin{figure*}[tbp]
\includegraphics[width=0.7\textwidth]{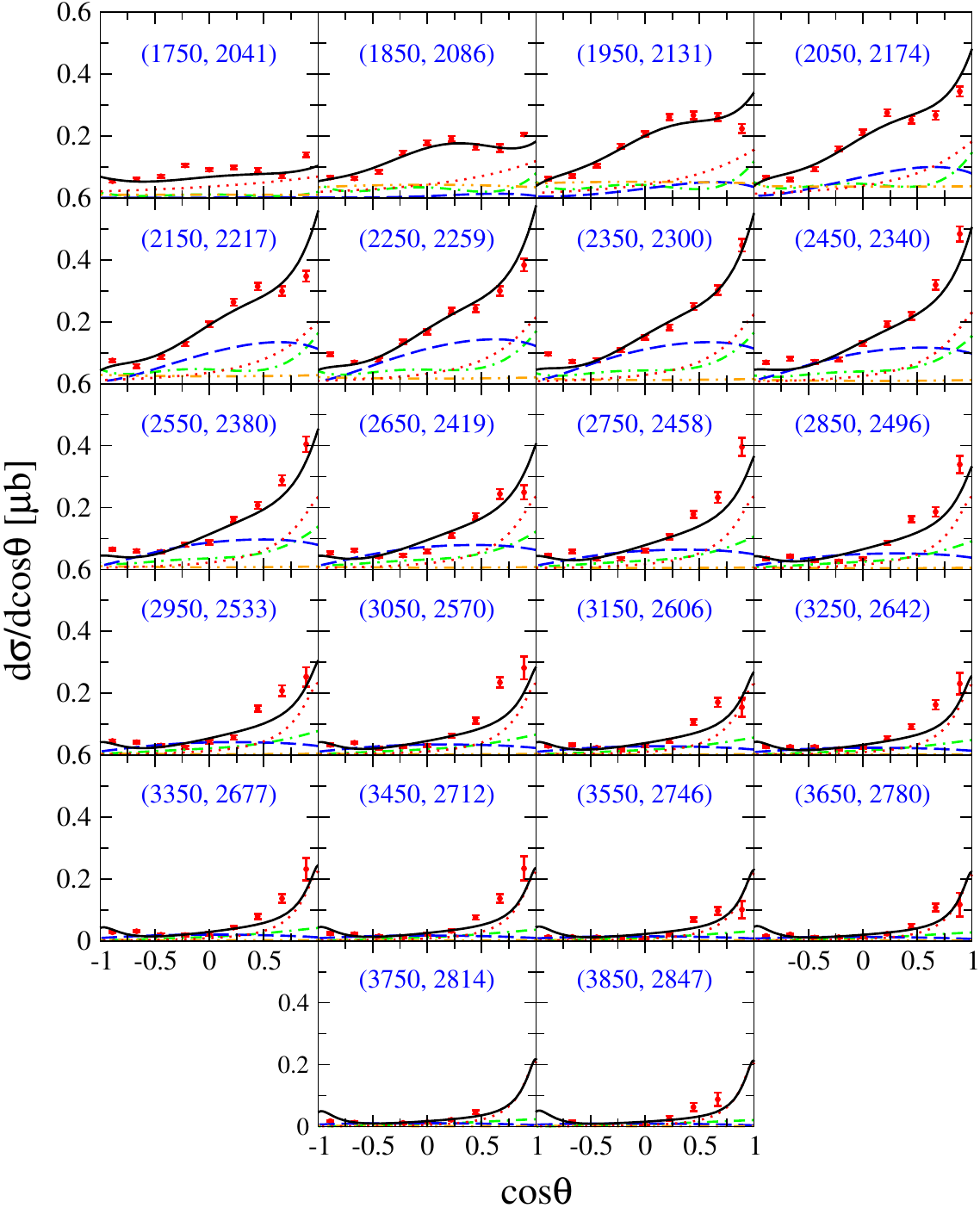}
\caption{Differential cross sections for $\gamma p\rightarrow K^{*+}\Lambda$ as a function of $\cos \theta$ in the center-of-mass frame (black solid line). The red dotted, blue dashed, green dash-dotted, and orange double-dot-dashed lines represent the individual contributions from the $K$, $N(2000)5/2^+$, $N(2060)5/2^-$, and $N(2080)3/2^-$ exchanges, respectively. Data are taken from the CLAS Collaboration \cite{CLAS:2013qgi}. The numbers in parentheses denote the photon laboratory incident energy (left number) and the total center-of-mass energy of the system (right number), both in MeV. }
\label{fig_dsig}
\end{figure*}

\begin{figure*}[tbp]
\includegraphics[width=0.7\textwidth]{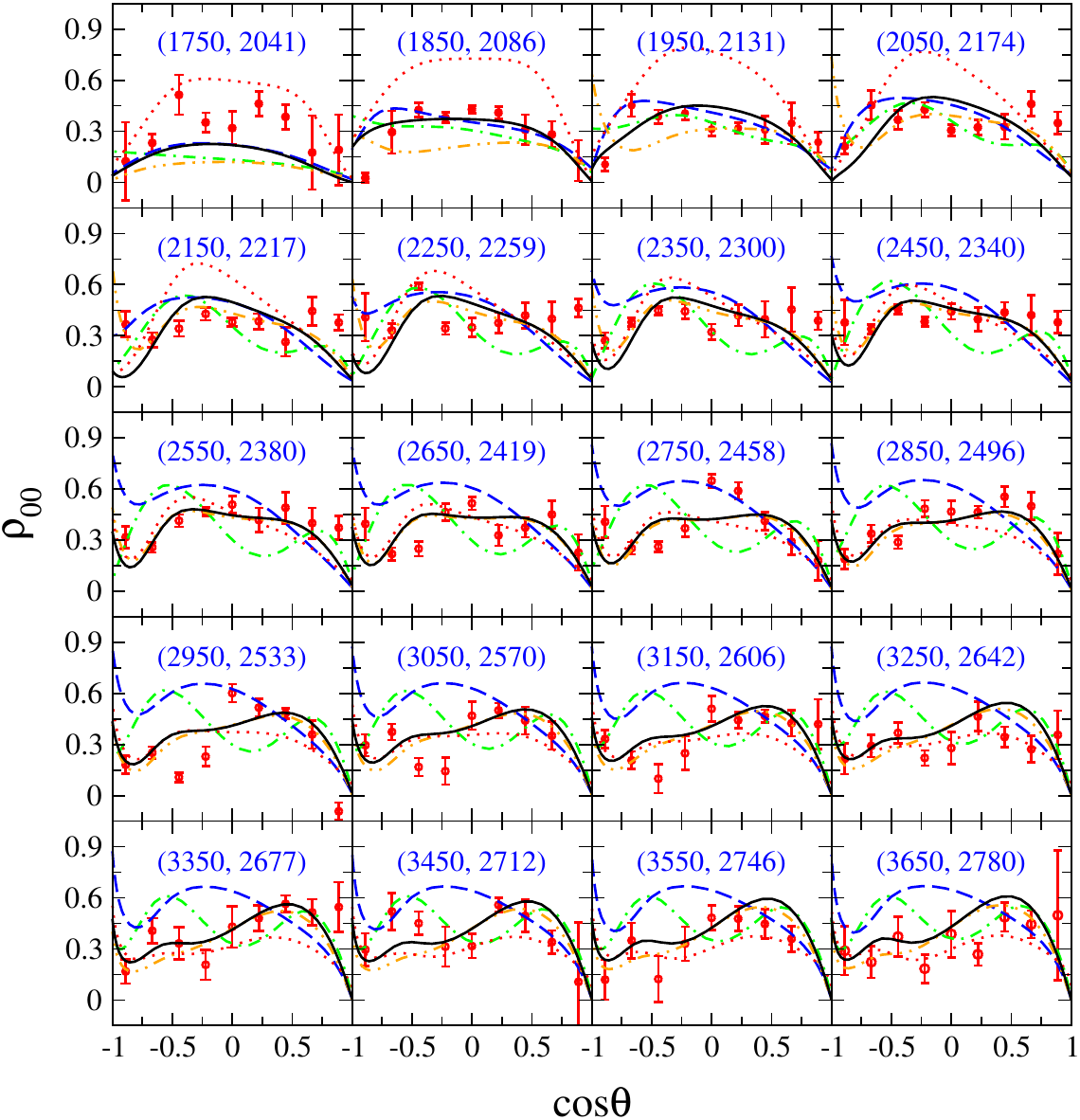}
\caption{Spin density matrix elements $\rho_{00}$ for $\gamma p\rightarrow K^{*+}\Lambda$ as a function of $\cos \theta$ in the center-of-mass frame (black solid line). The red dotted, blue dashed, green dash-dotted, and orange double-dot-dashed lines denote effects of the individual contributions from the $K$, $N(2000)5/2^+$, $N(2060)5/2^-$, and $N(2080)3/2^-$ exchanges, respectively, obtained by switching off the corresponding amplitude in the full reaction amplitude. Data are taken from the CLAS Collaboration \cite{CLAS:2017sgi}. The numbers in parentheses denote the photon laboratory incident energy (left number) and the total center-of-mass energy of the system (right number), both in MeV.}
\label{fig_rho00}
\end{figure*}

\begin{figure*}[tbp]
\includegraphics[width=0.7\textwidth]{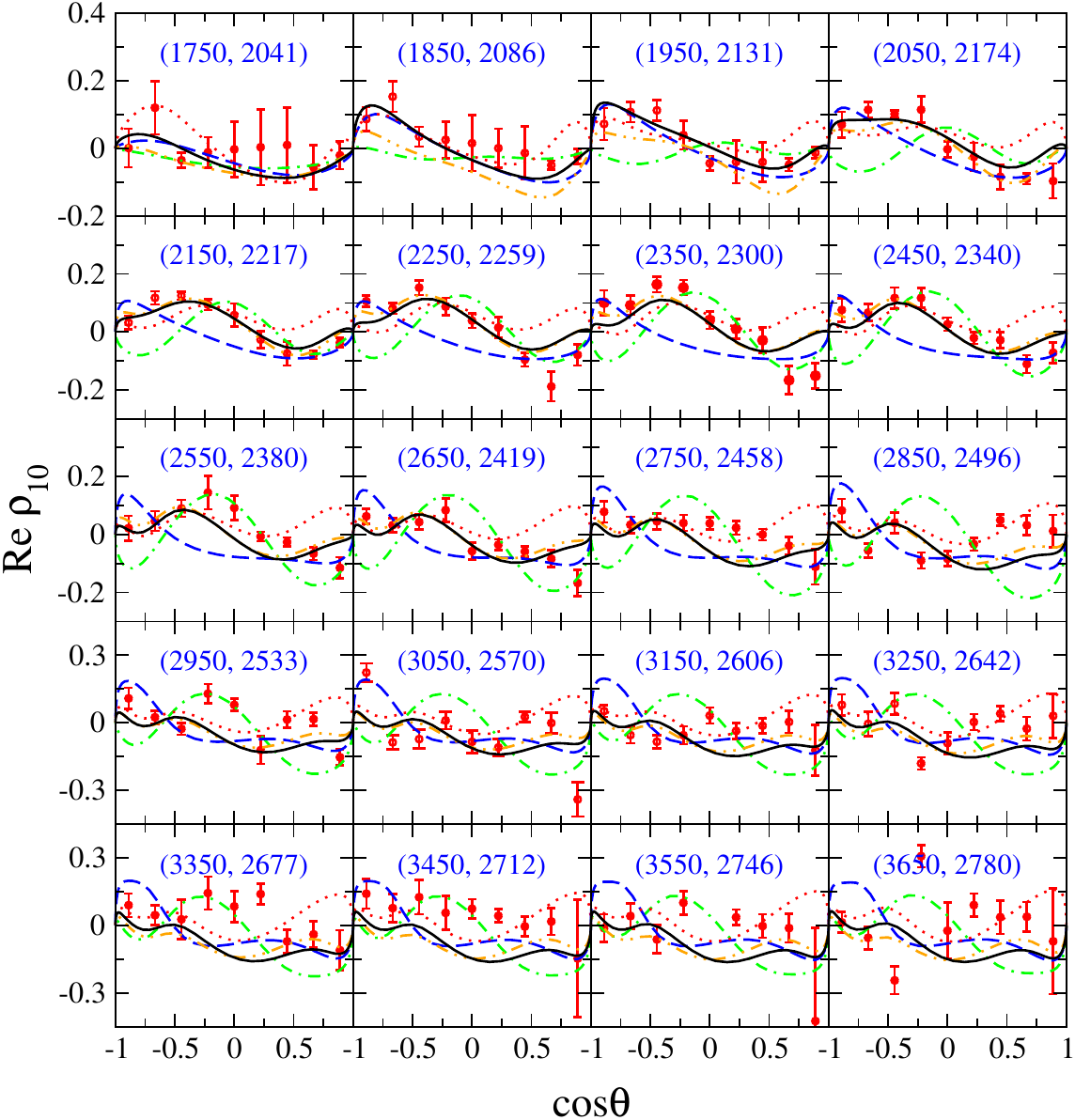}
\caption{Real part of the spin density matrix elements $\rho_{10}$ for $\gamma p\rightarrow K^{*+}\Lambda$ as a function of $\cos \theta$ in the center-of-mass frame (black solid line). Notations are the same as Fig.~\ref{fig_rho00}.}
\label{fig_rho10}
\end{figure*}

\begin{figure*}[tbp]
\includegraphics[width=0.7\textwidth]{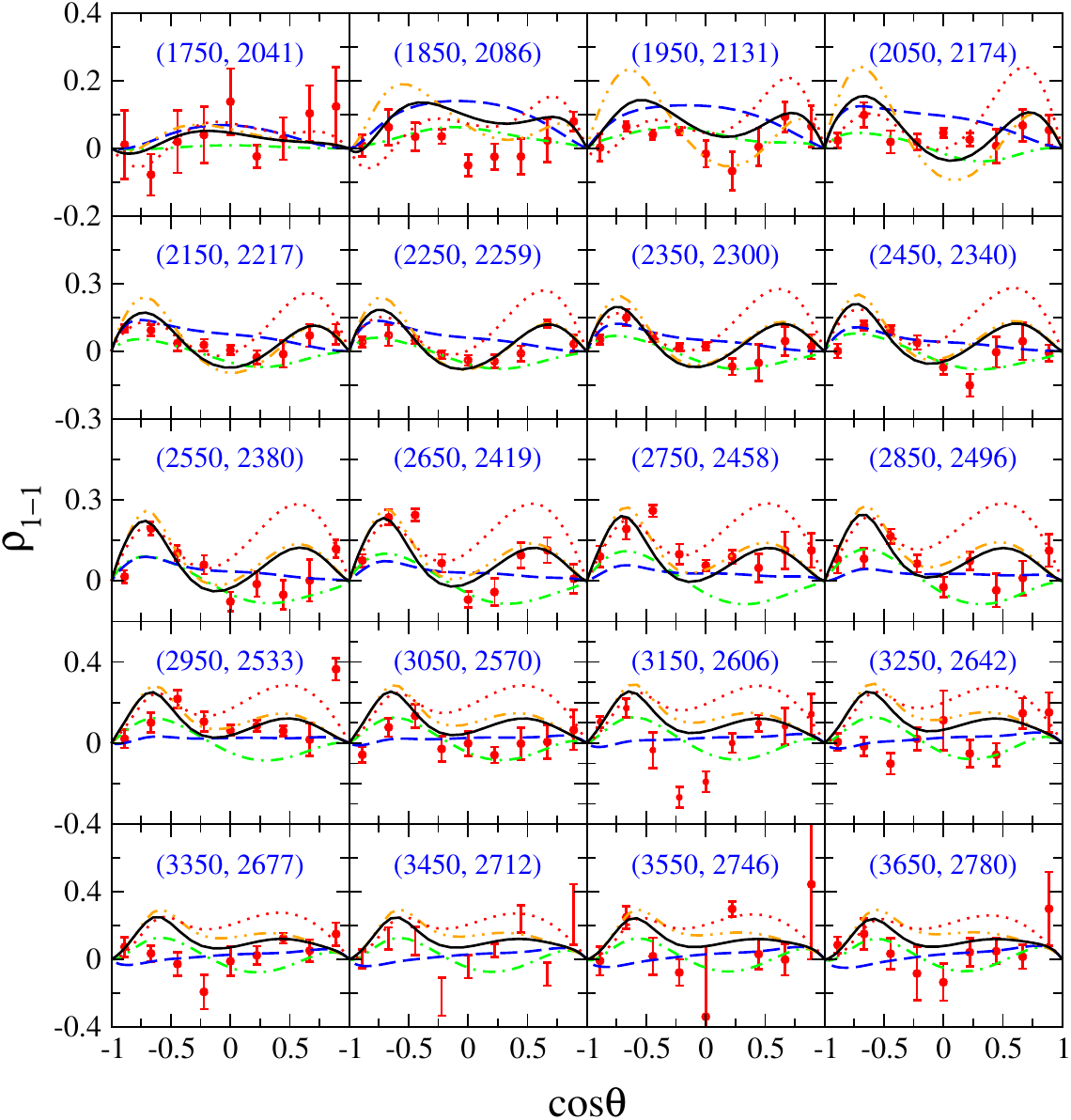}
\caption{Spin density matrix elements $\rho_{1-1}$ for $\gamma p\rightarrow K^{*+}\Lambda$ as a function of $\cos \theta$ in the center-of-mass frame (black solid line). Notations are the same as Fig.~\ref{fig_rho00}.}
\label{fig_rho1m1}
\end{figure*}

\section{Results and discussion}   \label{Sec:results}

Originally, we have studied the $\gamma p\to K^{*+}\Lambda$ reaction within an effective Lagrangian model in Ref.~\cite{Wang:2017tpe}. There, the high-precision cross-section data for $\gamma p\to K^{*+}\Lambda$ \cite{CLAS:2013qgi} have been analyzed, and it was found that apart from the $t$-channel $K$, $\kappa$, and $K^*$ exchanges, the $s$-channel $N$ exchange, the $u$-channel $\Lambda$, $\Sigma$, and $\Sigma^*(1385)$ exchanges, and the generalized contact term, the contributions from at least two near-threshold nucleon resonances in the $s$ channel should be considered in constructing the reaction amplitude to achieve a satisfactory description of the cross-section data. One of these required resonances is the $N(2060){5/2}^-$, which is responsible for the shape of the angular distributions near the $K^*\Lambda$ threshold, and the other needed resonance could be one of the $N(2000){5/2}^+$, $N(2040){3/2}^+$, $N(2100){1/2}^+$, $N(2120){3/2}^-$, and $N(2190){7/2}^-$ resonances. Further in Ref.~\cite{Wei:2020fmh}, we have reanalyzed the $\gamma p\to K^{*+}\Lambda$ reaction by incorporating  the later published data on spin density matrix elements for $K^\ast$ \cite{CLAS:2017sgi} into the data set. By doing so, more constraints have been imposed on the theoretical model, and it was found that with the new data on spin density matrix elements being taken into account, the $N(2000){5/2}^+$ resonance, rather than one of the $N(2040){3/2}^+$, $N(2100){1/2}^+$, $N(2120){3/2}^-$, and $N(2190){7/2}^-$ resonances, is necessary along side the $N(2060){5/2}^-$ resonance to achieve a satisfactory description of the data on both the differential cross sections and the spin density matrix elements for $\gamma p\to K^{*+}\Lambda$. The $t$-channel $K$ exchange and the $s$-channel $N(2060){5/2}^-$ and $N(2000){5/2}^+$ exchanges were found to make dominant contributions. 

Although our previous work of Ref.~\cite{Wei:2020fmh} provides so far the most recent and most comprehensive analysis of all available data on both differential cross sections and spin density matrix elements for the $\gamma p\to K^{*+}\Lambda$ reaction, there is a fly in the ointment: the data on spin density matrix elements $\rho_{00}$ in the center-of-mass energy region around $W\approx 2.08$ GeV are much underestimated despite that the overall agreement of the theoretical results with the data is rather satisfactory. This discrepancy, as mentioned previously, indicates that additional reaction mechanisms may be needed in our theoretical model. A natural candidate to address this issue is the $N(2080)3/2^-$, a $K^\ast \Sigma$ molecular state proposed as the strange partner of the $P_c^+(4457)$ \cite{Lin:2018kcc,He:2017aps,Zou:2018uji}, as its active role has already been demonstrated in the $K^{\ast +}\Sigma^0$, $K^{\ast 0}\Sigma^+$, and $\phi p$ photoproduction reactions \cite{Ben:2023uev,Wu:2023ywu}.

With the inclusion of the contributions from the $s$-channel $N(2080)3/2^-$ exchange to the effective Lagrangian model proposed in our previous works of Refs.~\cite{Wang:2017tpe,Wei:2020fmh}, we refit all available data on both differential cross sections and spin density matrix elements for the $\gamma p\to K^{*+}\Lambda$ reaction. The fitted model parameters are listed in Tables~\ref{tab:para1} and \ref{tab:para2}. Note that the mass and width of $N(2080)3/2^-$ in the present work are fixed at $2080$ MeV and $141$ MeV, respectively, as given in Ref.~\cite{Lin:2018kcc}. One can see from Tables~\ref{tab:para1} and \ref{tab:para2} that, after including the contributions from $N(2080)3/2^-$, the fitted masses, widths, and cutoff parameters for the $N(2060)5/2^-$ and $N(2000)5/2^+$ resonances are comparable to those obtained in Ref.~\cite{Wei:2020fmh}. Additionally, the value of the cutoff parameter for the $t$-channel $K$ exchange in the new fit is rather close to that obtained in Ref.~\cite{Wei:2020fmh}. This is not difficult to understand if we notice that, as found in Ref.~\cite{Wei:2020fmh}, the contributions of the $t$-channel $K$ exchange are primarily constrained by the differential cross-section data at forward angles in the high-energy region, while the contributions of the $s$-channel $N(2060)5/2^-$ and $N(2000)5/2^+$ exchanges are largely restricted by the shape of the angular distributions in the near-threshold energy region. Unsurprisingly, the $s$-channel $N(2080)3/2^-$ exchange is expected to provide complementary contributions to the differential cross sections while considerable contributions to the spin density matrix elements through its interference with other terms in the reaction amplitude.

To visually access the improvement or deterioration in the fitting quality for the considered data, in Table~\ref{table_chi}, the evaluated $\chi^2$ per number of data points, $\chi_i^2/N_i$, for a given type of observable specified by the index $i=d\sigma$ (differential cross section), $\rho_{00}$, ${\rm Re}\,\rho_{10}$, and $\rho_{1-1}$, with $N_i$ being the corresponding number of data points considered, are listed and compared to those obtained in Ref.~\cite{Wei:2020fmh}. There, the numbers in the brackets denote the corresponding values for the data in the energy range $W \le 2217$ MeV, while the last row corresponds to the global $\chi^2/N$, with $N$ being the total number of data points considered. One can see that, overall, the inclusion of $N(2080)3/2^-$ can significantly improve the fit quality of the data for $\gamma p \to K^{*+}\Lambda$, especially in the near-threshold energy region with $W\le 2217$ MeV. Specifically, the inclusion of the $N(2080)3/2^-$ has leaded to a better description of the data on $\rho_{00}$ and $\rho_{1-1}$, and a slightly worse description of the data on ${\rm Re}\,\rho_{10}$ and differential cross sections. Focusing on the near-threshold energy region ($W\le 2217$ MeV) where the contributions of the $N(2080)3/2^-$ are most relevant, one sees that the inclusion of the $N(2080)3/2^-$ has helped to provide an overwhelmingly better description of the data on spin density matrix elements $\rho_{00}$. In addition, it has resulted in a much better description of the ${\rm Re}\,\rho_{10}$ and $\rho_{1-1}$, while maintaining same quality description of the differential cross section data. A more detailed analysis of the effects of the $N(2080)3/2^-$ will be presented below. 

The results for the differential cross sections ${\rm d}\sigma/{\rm d}\cos\theta$ and the spin density matrix elements $\rho_{00}$, ${\rm Re}\,\rho_{10}$, and $\rho_{1-1}$ calculated by use of the parameters listed in Tables~\ref{tab:para1} and \ref{tab:para2} are presented in Fig.~\ref{fig_dsig} and Figs.~\ref{fig_rho00}-\ref{fig_rho1m1}, respectively. There, the numbers in parentheses denote the photon laboratory incident energy (left number) and the total center-of-mass energy of the system (right number), both in MeV.

In Fig.~\ref{fig_dsig}, the black solid lines denote the differential cross sections as a function of $\cos \theta$ in the center-of-mass frame, calculated from the full reaction amplitude. The red dotted, blue dashed, green dash-dotted, and orange double-dot-dashed lines represent the individual contributions from the $K$, $N(2000)5/2^+$, $N(2060)5/2^-$, and $N(2080)3/2^-$ exchanges, respectively. Contributions from the other terms are too small to be clearly seen with the scale used and are therefore not shown in Fig.~\ref{fig_dsig}.  One can see from Fig.~\ref{fig_dsig} that the contributions from the $K$ exchange are significant across the entire energy region considered, and especially, they dominate the angular distributions at forward angles in the high-energy region. The exchanges of $N(2060)5/2^-$, $N(2000)5/2^+$, and $N(2080)3/2^-$ contribute mainly in the low-energy region, responsible for the shapes of the angular distributions. Compared to the cross-section results in Fig.~5 of Ref.~\cite{Wei:2020fmh}, the individual contributions of the $K$, $N(2060)5/2^-$, and $N(2000)5/2^+$ exchanges in the present work remain almost unchanged but are slightly smaller. This observation is consistent with the fact that the parameters for the $K$, $N(2060)5/2^-$, and $N(2000)5/2^+$ exchanges determined in the present work, as listed in Tables~\ref{tab:para1} and \ref{tab:para2}, are close to those determined in Ref.~\cite{Wei:2020fmh}.

In Figs.~\ref{fig_rho00}-\ref{fig_rho1m1}, the black solid lines represent the spin density matrix elements $\rho_{00}$, ${\rm Re}\,\rho_{10}$, and $\rho_{1-1}$, respectively, as functions of $\cos \theta$ in the center-of-mass frame, calculated from the full reaction amplitude. The red dotted, blue dashed, green dash-dotted, and orange double-dot-dashed lines denote the effects of the individual contributions from the $K$, $N(2000)5/2^+$, $N(2060)5/2^-$, and $N(2080)3/2^-$ exchanges, respectively, obtained by switching off the corresponding amplitude in the full reaction amplitude. Effects from other terms on $\rho_{00}$, ${\rm Re}\,\rho_{10}$, and $\rho_{1-1}$ are relatively small and are therefore not shown in these figures. From Figs.~\ref{fig_rho00}-\ref{fig_rho1m1}, it is evident that, overall, the theoretical results for $\rho_{00}$, ${\rm Re}\,\rho_{10}$, and $\rho_{1-1}$ agree satisfactorily well with the corresponding data. The $K$, $N(2000)5/2^+$, and $N(2060)5/2^-$ exchanges have significant effects on $\rho_{00}$, ${\rm Re}\,\rho_{10}$, and $\rho_{1-1}$ across the entire energy region considered. The contributions from the $N(2000)5/2^+$ and $N(2060)5/2^-$ exchanges in the high-energy region are mainly due to their interference with the $K$-meson exchange. Similar to the situation with the differential cross sections, the $N(2080)3/2^-$ exchange has noticeable effects mainly in the low-energy region, indicating that its interference with the $K$ exchange in the high-energy region is relatively weak, in contrast to those of the $N(2000)5/2^+$ and $N(2060)5/2^-$ resonances. Additionally, it is observed that the effects of the $N(2080)3/2^-$ exchange on $\rho_{00}$ are much stronger than on ${\rm Re}\,\rho_{10}$ and $\rho_{1-1}$. Actually, it is the inclusion of the $N(2080)3/2^-$ exchange that makes the distinct improvement in the theoretical description of the $\rho_{00}$ data in the near-threshold energy region. Compared to the results for the spin density matrix elements in Figs.~6-8 of Ref.~\cite{Wei:2020fmh}, the main features of the effects of the $K$, $N(2000)5/2^+$, and $N(2060)5/2^-$ exchanges in the present work remain roughly unchanged but are slightly softer, leaving some room for the contributions of the $N(2080)3/2^-$ exchange. In terms of fitting quality, visually the agreement of the theoretical results for $\rho_{00}$ and the corresponding data in the near-threshold energy region is overwhelmingly better in the present work than in Ref.~\cite{Wei:2020fmh}, especially in the lowest two energy bins, with the credit going to the newly considered $N(2080)3/2^-$ exchange in the present work. For ${\rm Re}\,\rho_{10}$ and $\rho_{1-1}$, the results in the present work are visually similar to those from Ref.~\cite{Wei:2020fmh}.

\begin{figure}[tbp]
\includegraphics[width=\columnwidth]{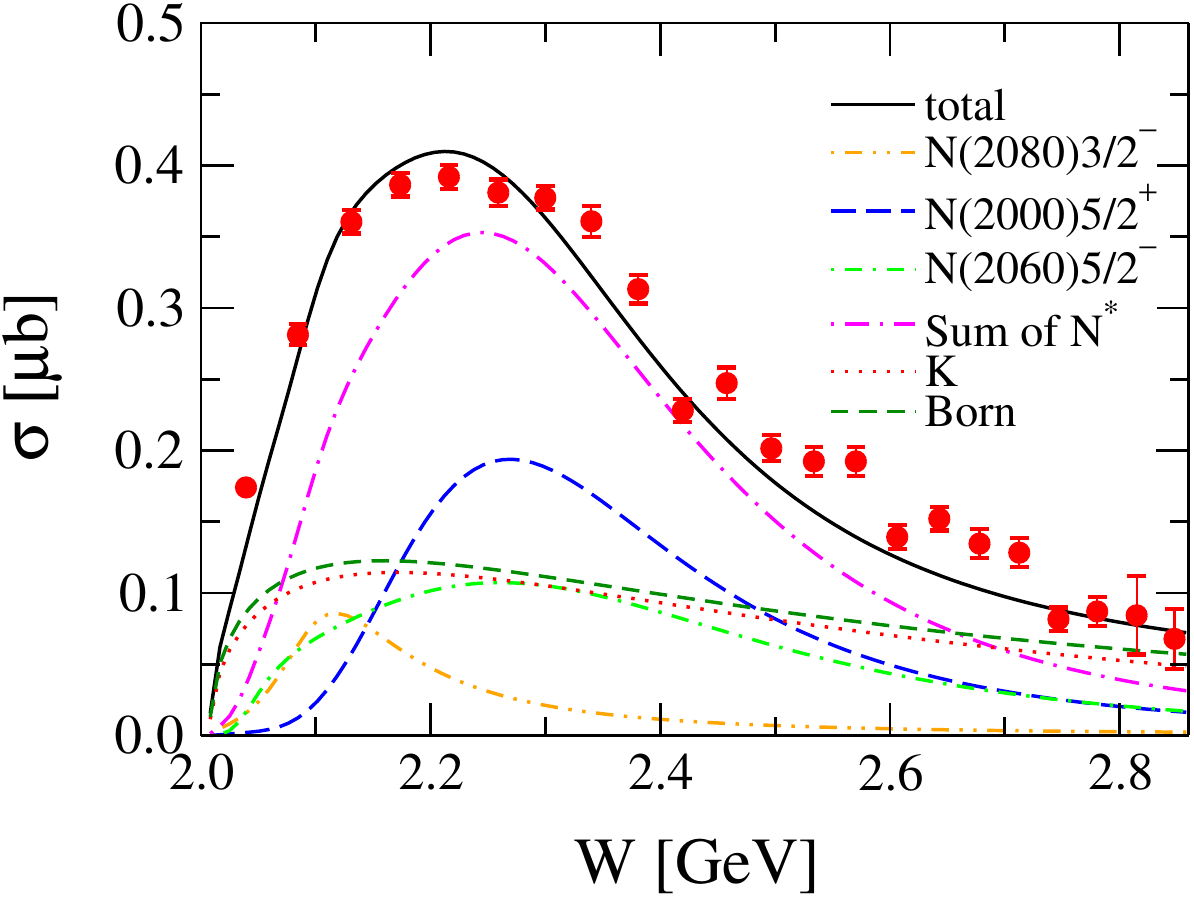}
\caption{Total cross sections with individual contributions for $\gamma p\rightarrow K^{*+}\Lambda$ as a function of the center-of-mass energy of the system. The data are taken from the CLAS Collaboration \cite{CLAS:2013qgi} but not included in the fitting procedure. }
\label{fig_cross}
\end{figure}

The total cross sections for $\gamma p \to K^{*+}\Lambda$ predicted in the present work are shown in Fig.~\ref{fig_cross}, where the contributions from various important individual terms are also displayed. Consistent with the situation of the differential cross sections, the model results can describe the total cross-section data very well. Note that these total cross-section data are not included in the fitting procedure. One sees from Fig.~\ref{fig_cross} that the contributions from the $N(2000)5/2^+$, $N(2060)5/2^-$, and $N(2080)3/2^-$ exchanges are significant, with the sum of them dominating the cross sections in the low-energy region and responsible for the bump structure observed in the data around the center-of-mass energy $W\approx 2.2$ GeV. The $K$ exchange also provides significant contributions across the entire energy region considered, and in particular, it becomes more and more prominent as the energy increases. Compared to the total cross-section results of Ref.~\cite{Wei:2020fmh}, the major difference is that in the near-threshold energy region, the combined contributions from the resonances in the present work are noticeably larger than those in Ref.~\cite{Wei:2020fmh}, due to the inclusion of the newly considered $N(2080)3/2^-$ exchange. Correspondingly, the contributions from the $K$ exchange in the present work are smaller than those in Ref.~\cite{Wei:2020fmh}, as the fitted value of the cutoff parameter $\Lambda_K$ is smaller in the present work.

From all the results presented above, it is evident that by incorporating the $N(2080)3/2^-$ exchange into our effective Lagrangian model for $\gamma p \to K^{*+}\Lambda$, the theoretical description of the data on differential cross sections and spin density matrix elements $\rho_{00}$ in the near-threshold energy region is significantly improved. In particular, the results for $\rho_{00}$ in the lowest two energy bins are now quite well described, effectively resolving the issue of the large underestimation of these data in Ref.~\cite{Wei:2020fmh}. This achievement indicates that the hypothesis of the $N(2080)3/2^-$ as a $K^\ast\Sigma$ molecular state is compatible with the available data for the $\gamma p \to K^{*+}\Lambda$ reaction. It is worth noting that similar findings have also been reported in theoretical investigations of the $K^{*+}\Sigma^0$,  $K^{*0}\Sigma^+$, and $\phi p$ photoproduction reactions \cite{Ben:2023uev,Wu:2023ywu}.

Finally, we note that the exchange of the $N(2270)3/2^-$, which was proposed as a molecular state composed of $K^\ast\Sigma^\ast$ whose threshold lies at $2277$ MeV and has been shown to play significant roles in the $K^{*+}\Sigma^0$,  $K^{*0}\Sigma^+$, and $\phi p$ photoproduction reactions \cite{Ben:2023uev,Wu:2023ywu}, is not included in the present study of the $\gamma p \to K^{*+}\Lambda$ reaction. 
Actually, we have tested replacing either the $N(2000)5/2^+$ or $N(2060)5/2^-$ resonance with the $N(2270)3/2^-$ molecular state in our model. However, this substitution resulted in a noticeable deterioration of the fit quality. Incorporating the contributions of $N(2270)3/2^-$ as an additional interaction mechanism alongside the $N(2000)5/2^+$, $N(2060)5/2^-$, and $N(2080)3/2^-$ exchanges could improve the agreement between the model results and the data, since this would introduce additional model parameters. We postpone such refinements until more data become available for the $\gamma p \to K^{*+}\Lambda$ reaction in the future.

As a final remark, note that besides the direct calculations of these triangle diagrams using the LoopTools package, one can also partially mimic the corresponding loop diagrams in Fig.~\ref{fig:loop} by a phase factor $\exp[i\phi_{R}]$ multiplied to the amplitude of the corresponding resonance exchange with $\phi_{R}$ being a fit parameter, as done in Refs.~\cite{Ben:2023uev,Wu:2023ywu}. To test the effects of different parametrizations, we have also performed this method, with the results presented in Table~\ref{table_chi_phase}. Actually, in this alternative parametrization we need to introduce four additional parameters compared to the two-resonance case~\cite{Wei:2020fmh}. The triangle loop contribution is represented by $\exp[i\phi_{R}]$ herein. Although the resulting $\chi^2/N$ is slightly smaller than that in our main analysis (cf. Table~\ref{table_chi}), it yields a much worse description of $\rho_{1-1}$ in the lower-energy region. In both cases, the $N(2000)5/2^+$, $N(2060)5/2^-$ resonances and the $K$-exchange play important roles and yield the same descriptive tendencies for the observables. In general, the two parametrizations do not differ significantly to each other in the quality of the fit. We choose the explicit triangle loops as our main parametrization, since they exhibit a clearer physical picture for a hadronic molecule.

\begin{table}[tbp]
    \caption{The values of $\chi_i^2/N_i$ obtained from the phase parametrization. Notations are the same as in Table~\ref{table_chi}.}
\begin{tabular*}{\columnwidth}{@{\extracolsep\fill}ccccc}
    \hline\hline
$\chi_{d\sigma}^2/N_{d\sigma}$ &$\chi_{\rho_{00}}^2/N_{\rho_{00}}$ &$\chi_{{\rm Re}\,\rho_{10}}^2/N_{{\rm Re}\,\rho_{10}}$ &$\chi_{\rho_{1-1}}^2/N_{\rho_{1-1}}$ &$\chi^2/N$ \\\hline
 $2.98~(0.83)$  &   $5.55~(4.66)$  &  $5.33~(2.42)$  &  $5.85~(7.42)$  &   $4.90~(3.83)$   \\ 
  \hline\hline
\end{tabular*}
\label{table_chi_phase}
\end{table}

\section{Summary}   \label{sec:summary}

In our previous work of Ref.~\cite{Wei:2020fmh}, we have analyzed the available data on differential cross sections \cite{CLAS:2013qgi} and spin density matrix elements \cite{CLAS:2017sgi} for the $\gamma p\to K^{\ast +}\Lambda$ reaction within an effective Lagrangian approach. There, the $t$-channel $K$, $K^*$, and $\kappa$ exchanges, the $u$-channel $\Lambda$, $\Sigma$, and $\Sigma^*$ exchanges, the $s$-channel $N$, $N(2060)5/2^-$, and $N(2000)5/2^+$ exchanges, and the interaction current were taken into account in constructing the reaction amplitudes. It was found that the overall agreement of the theoretical results with the corresponding data is satisfactory, with the dominant contributions arising from the $K$, $N(2060)5/2^-$, and $N(2000)5/2^+$ exchanges. Nevertheless, the model significantly underestimates the data for the spin density matrix elements $\rho_{00}$ in the center-of-mass energy region around $W\approx 2.08$ GeV. This discrepancy suggests that additional interaction mechanisms are needed to improve the description of the data for $\gamma p\to K^{\ast +}\Lambda$ in our model proposed in Ref.~\cite{Wei:2020fmh}.  

The $P_c^+(4457)$ state observed by the LHCb Collaboration \cite{LHCb:2019kea} has sparked extensive discussions regarding its structure and decay properties. One plausible interpretation is that it is a hadronic molecule composed of ${\bar D}^\ast \Sigma_c$. This interpretation is rather natural, as the reported mass of the $P_c^+(4457)$ state is approximately $2$ MeV below the ${\bar D}^\ast \Sigma_c$ threshold, which lies at $4459$ MeV. In the light quark sector, the $N(2080)3/2^-$ state has been proposed to be the strange partner of the $P_c^+(4457)$ molecule \cite{Lin:2018kcc,He:2017aps,Zou:2018uji}. Analogously, it is assumed to be a hadronic molecular state composed of $K^\ast \Sigma$ whose threshold is at $2086$ MeV. 

In the present work, we investigate the effects of the $N(2080)3/2^-$ molecular state in the $\gamma p \to K^{*+}\Lambda$ reaction. Assuming that the $N(2080)3/2^-$ exchange represents the interaction mechanism missed in our previously proposed model \cite{Wei:2020fmh}, we reanalyze all available data on both differential cross sections and spin density matrix elements for the $\gamma p \to K^{*+}\Lambda$ reaction. The results demonstrate that with the inclusion of the $N(2080)3/2^-$ exchange in our model, the overall agreement of the theoretical results with the corresponding data is significantly improved. In particular, the data on spin density matrix elements $\rho_{00}$ and ${\rm Re}\,\rho_{10}$ in the near-threshold energy region are now considerably better described, perfectly solving the issue of pronounced discrepancies between the model results of $\rho_{00}$ and the corresponding data in the lowest two energy bins in Ref.~\cite{Wei:2020fmh} [see illustration in Fig.~\ref{fig_wei}]. This achievement indicates the compatibility of the hadronic molecular picture of $N(2080)3/2^-$ with the available data for the $\gamma p \to K^{*+}\Lambda$ photoproduction reaction.

To further elucidate the reaction mechanisms for $\gamma p \to K^{*+}\Lambda$, contributions from various individual terms that compose the full reaction amplitude are analyzed. It is found that the contributions from the $N(2000)5/2^+$, $N(2060)5/2^-$, and $N(2080)3/2^-$ exchanges are dominant in the low-energy region, while the $K$ exchange contributes significantly across the entire energy region considered and becomes particularly prominent at forward angles in the higher-energy region.

In the future, this study can be developed in various directions. For example, as already mentioned, the inclusion of $N(2270)$ is feasible once more experimental data become available. Moreover, the current model is still based on individual tree-level and one-loop Feynman diagrams, without the two-body unitarity of $K^*\Lambda$. As the next step, unitarization approaches can be performed so that more precise dynamics can be embedded, which will likely improve the description of the data significantly. In addition, unitarity allows the rigorous definition of the resonances as the poles on the second Riemann sheet.

\begin{acknowledgments}
This work is partially supported by the National Natural Science Foundation of China under Grants No.~12305137, No.~12175240, and No.~12070131001, and the Fundamental Research Funds for the Central Universities.
\end{acknowledgments}

\end{document}